\documentclass[useAMS,usenatbib]{mn2e}
\usepackage{amsmath}
\usepackage{graphicx}
\usepackage{amssymb}
\usepackage{sidecap}
\usepackage{color}
\usepackage{enumerate}
\usepackage{algorithm}
\usepackage{algpseudocode}
\usepackage{verbatim}
\usepackage{changepage}

\def\dalemb#1#2{{\vbox{\hrule height.#2pt
        \hbox{\vrule width.#2pt height#1pt \kern#1pt \vrule width.#2pt}
        \hrule height.#2pt}}}

\def\ba{\begin{eqnarray}}
\def\ea{\end{eqnarray}}
\def\be{\begin{equation}}
\def\ee{\end{equation}}

\def\gtorder{\mathrel{\raise.3ex\hbox{$>$}\mkern-14mu
             \lower0.6ex\hbox{$\sim$}}}
\def\ltorder{\mathrel{\raise.3ex\hbox{$<$}\mkern-14mu
             \lower0.6ex\hbox{$\sim$}}}

\def\be{\begin{equation}}
\def\ee{\end{equation}}

\def\gtorder{\mathrel{\raise.3ex\hbox{$>$}\mkern-14mu
             \lower0.6ex\hbox{$\sim$}}}
\def\ltorder{\mathrel{\raise.3ex\hbox{$<$}\mkern-14mu
             \lower0.6ex\hbox{$\sim$}}}

\title[ACT: radio galaxy bias through cross-correlation with lensing]{The Atacama Cosmology Telescope: measuring radio galaxy bias through cross-correlation with lensing}

\author[R.~Allison et al.]{\parbox{\textwidth}{
Rupert Allison$^1$\thanks{E-mail:~\texttt{rupert.allison@astro.ox.ac.uk}},
Sam N. Lindsay$^1$,
Blake D. Sherwin$^2$,
Francesco de Bernardis$^3$,
J. Richard Bond$^4$,
Erminia Calabrese$^1$,
Mark J. Devlin$^5$,
Joanna Dunkley$^1$,
Patricio Gallardo$^3$,
Shawn Henderson$^3$,
Adam D. Hincks$^6$,
Ren\'{e}e Hlozek$^7$,
Matt Jarvis$^{1,8}$,
Arthur Kosowsky$^9$,
Thibaut Louis$^1$,
Mathew Madhavacheril$^{10}$,
Jeff McMahon$^{11}$,
Kavilan Moodley$^{12}$,
Sigurd Naess$^1$,
Laura Newburgh$^{13}$,
Michael D. Niemack$^3$,
Lyman A. Page$^{14}$,
Bruce Partridge$^{15}$,
Neelima Sehgal$^{10}$,
David N. Spergel$^7$,
Suzanne T. Staggs$^{14}$,
Alexander van Engelen$^{4}$,
Edward J. Wollack$^{16}$
}
\vspace{0.5cm}\\
\parbox{\textwidth}{
$^1$Sub-department of Astrophysics, University of Oxford, Denys Wilkinson Building, Oxford, OX1 3RH, UK \\
$^2$Berkeley Center for Cosmological Physics, LBL and Department of Physics, University of California, Berkeley, CA, USA 94720 \\
$^3$Department of Physics, Cornell University, Ithaca, NY, USA 14853 \\
$^4$Canadian Institute for Theoretical Astrophysics, University of Toronto, Toronto, ON, Canada M5S 3H8 \\
$^5$Department of Physics and Astronomy, University of Pennsylvania, 209 South 33rd Street, Philadelphia, PA, USA 19104 \\
$^6$Department of Physics and Astronomy, University of British Columbia, Vancouver, BC, Canada V6T 1Z4 \\
$^7$Department of Astrophysical Sciences, Peyton Hall, Princeton University, Princeton, NJ USA 08544 \\
$^8$Physics Department, University of the Western Cape, Bellville 7535, South Africa \\
$^9$Department of Physics and Astronomy, University of Pittsburgh, Pittsburgh, PA, USA 15260 \\
$^{10}$Physics and Astronomy Department, Stony Brook University, Stony Brook, NY USA 11794 \\
$^{11}$Department of Physics, University of Michigan, Ann Arbor, USA 48103 \\
$^{12}$Astrophysics and Cosmology Research Unit, School of Mathematics, Statistics and Computer Science, University of KwaZulu-Natal, Durban 4041, South Africa \\
$^{13}$Dunlap Institute for Astronomy and Astrophysics, University of Toronto, 50 St. George St., Toronto ON, Canada M5S 3H4 \\
$^{14}$Joseph Henry Laboratories of Physics, Jadwin Hall, Princeton University, Princeton, NJ, USA 08544 \\
$^{15}$Department of Physics and Astronomy, Haverford College, Haverford, PA, USA 19041 \\
$^{16}$NASA/Goddard Space Flight Center, Greenbelt, MD, USA 20771 \\
\vspace{-8mm}
}}
\begin{document}

\date{}

\pagerange{\pageref{firstpage}--\pageref{lastpage}} \pubyear{2014}
\maketitle
\begin{abstract} 
We correlate the positions of radio galaxies in the FIRST survey with the CMB lensing convergence estimated from the Atacama Cosmology Telescope over 470 deg$^2$ to determine the bias of these galaxies. We remove optically cross-matched sources below redshift $z=0.2$ to preferentially select Active Galactic Nuclei (AGN). We measure the angular cross-power spectrum $C_l^{\kappa g}$ at $4.4\sigma$ significance in the multipole range $100<l<3000$, corresponding to physical scales between $\approx$~2--60 Mpc at an effective redshift $z_{\rm eff}= 1.5$. Modelling the AGN population with a redshift-dependent bias, the cross-spectrum is well fit by the {\sl Planck} best-fit $\Lambda$CDM cosmological model. Fixing the cosmology we fit for the overall bias model normalization, finding $b(z_{\rm eff}) = 3.5 \pm 0.8$  for the full galaxy sample, and $b(z_{\rm eff})=4.0\pm1.1~(3.0\pm1.1)$ for sources brighter (fainter) than 2.5~mJy. This measurement characterizes the typical halo mass of radio-loud AGN: we find $\log(M_{\rm halo} / M_\odot) = 13.6^{+0.3}_{-0.4}$.
\end{abstract}

\label{firstpage}
\section{Introduction}
\label{intro}
Radio galaxies trace the large-scale structure in the Universe which has been measured with large-area surveys including FIRST, WENSS, NVSS, and SUMSS \citep{Becker:1995,Rengelink:1997,Condon:1998,Bock:1999}; for an overview see \citet{deZotti:2010}. The angular clustering of these galaxies has been measured by \citet{Cress:1996,Magliocchetti:1998,Blake:2002,Overzier:2003,Blake:2004,Lindsay:2014a}. The clustering of radio galaxies will soon be measured over much larger volumes of the universe with the Square Kilometer Array (SKA) and its precursors, allowing cosmological effects such as dark energy, modified gravity and non-Gaussianity to be probed \citep[e.g.,][]{Carilli:2004, Blake:2004b,Raccanelli:2012,Camera:2012,Maartens:2013,Norris:2013,Jarvis:2015}. 

The bias $b$ of a large-scale structure tracer relates overdensities of that tracer $\delta$ to overdensities of the underlying dark matter field $\delta_{\rm DM}$:
\begin{equation}
\delta = b \delta_{\rm DM}. 
\end{equation}
Radio-selected galaxies broadly contain two populations: high-redshift Active Galactic Nuclei (AGN) and low-redshift star-forming galaxies \citep{Condon:2002}. AGN dominate the radio emission at high flux ($\gtrsim 1$~mJy) and are highly biased, their hosts being among the most massive galaxies in the early universe \citep[e.g.,][]{Jarvis:2001b, Rocca:2004, Seymour:2007, deZotti:2010, Fernandes:2015}. Their bias depends strongly on galaxy mass and redshift \citep[e.g.,][]{Seljak:2004}, and is poorly constrained particularly at high redshift where few optical counterparts are observed. Some progress has been made by identifying redshifts spectroscopically: using Galaxy And Mass Assembly (GAMA) data the bias of FIRST radio galaxies was measured at $z \approx 0.34$ over 200 deg$^2$ to the 10\% level \citep{Lindsay:2014a}. On a smaller square degree region, clustering measurements using data from the Very Large Array (VLA) and VISTA Deep Extragalactic Observations \citep[VIDEO;][]{Jarvis:2013} were used to show evidence for a strongly increasing bias at $z>2$ \citep{Lindsay:2014b}.

An alternative way to constrain bias is through cross-correlation of the tracer fluctuations with gravitational lensing due to large-scale structure. In particular, the lensing of the Cosmic Microwave Background (CMB) measures the integrated matter fluctuations to $z \approx 1100$. As we will show the high-redshift radio source distribution overlaps strongly with the broad CMB lensing kernel. Cross-correlations between the CMB and other tracers of large-scale structure have been reported by e.g., \cite{Smith:2007, Hirata:2008, Feng:2012b, Bleem:2012, PlanckXVIII,vanEngelen:2014,Fornengo:2014}. The \citet{PlanckXVII} detect the correlation of lensing with radio galaxies from NVSS at $20 \sigma$. \cite{Sherwin:2012} correlate lensing measurements from the Atacama Cosmology Telescope (ACT) with optically-selected quasars from the Sloan Digital Sky Survey (SDSS), measuring a bias of b = $2.5 \pm 0.6$ at an effective redshift $z \approx 1.4$. \cite{Geach:2013} correlate lensing from the South Pole Telescope (SPT) with quasars selected from the Wide-field Infrared Survey Explorer (WISE), measuring a bias $b = 1.61 \pm 0.22$ at $z \approx 1.0$. An advantage of cross-correlations is that they are robust to systematic biases which may be particular to each dataset.

In this paper we measure the angular cross-power spectrum $C_l^{\kappa g}$ between the lensing convergence estimated from ACT with the FIRST radio source overdensity. We use lensing maps from the three-year ACT Equatorial survey \citep{Das:2013} together with the first-season ACTPol survey \citep{Naess:2014,Madhavacheril:2014,vanEngelen:2014}. We consider 36,000 radio sources with flux brighter than 1~mJy, and remove optically cross-matched sources from the Sloan Digital Sky Survey \citep{York:2000} at $z<0.2$ to preferentially select AGN, discarding the majority of low-redshift star-forming galaxies. We use this to estimate the bias normalization, assuming a fixed cosmological model, and using a redshift distribution and bias-evolution model from the simulated radio catalogue of the SKA Design Study \citep[SKADS,][]{Wilman:2008}. We measure $C_l^{\kappa g}$ across a wide range of scales ($100<l<3000$) and consider various splits of the radio sources to investigate redshift and flux dependence of the bias. 

We describe the lensing and radio data and the cross-correlation analysis methods in Section~\ref{sData}. The results and discussion are presented in Section~\ref{sResults}, with further interpretation of the AGN bias in Section~\ref{sBias}. We conclude in Section~\ref{sConclusion}. 

\section{Data and analysis}
\label{sData}
\subsection{ACT and ACTPol}
\label{ssACT}
The Atacama Cosmology Telescope (ACT) is located at an altitude of 5190m in Parque Astron\'o{}mico Atacama in Northern Chile. The telescope and its current polarization-sensitive receiver, ACTPol, are described in \cite{Niemack:2010}. The two seasons of ACT temperature data and the ACTPol first-season temperature and polarization data used in this analysis are presented in \cite{Das:2013} and \cite{Naess:2014}. Lensing by large-scale structure induces coupling of otherwise independent temperature and polarization modes. We construct estimators of the lensing convergence from quadratic combinations of temperature and polarization maps in Fourier space, following the methodology of \cite{Hu:2002}. We use the same lensing convergence maps and Monte-Carlo simulations described by \cite{Das:2011} and \cite{vanEngelen:2014}. 

In this analysis we use two ACT datasets. The first is the ACT Equatorial data which spans a thin strip along the celestial equator with an area of 300 deg$^2$. This strip is partitioned into six approximately equal area patches over which we compute the cross-spectrum separately then average (weighting by patch area) for the final result. We lose negligible information at the scales of interest and it allows for patch to patch consistency checks. The effective white-noise component of the two-season co-added data is $18~\mu{\rm K}$-arcmin. 

\begin{figure}
\centering
\includegraphics[width=80mm]{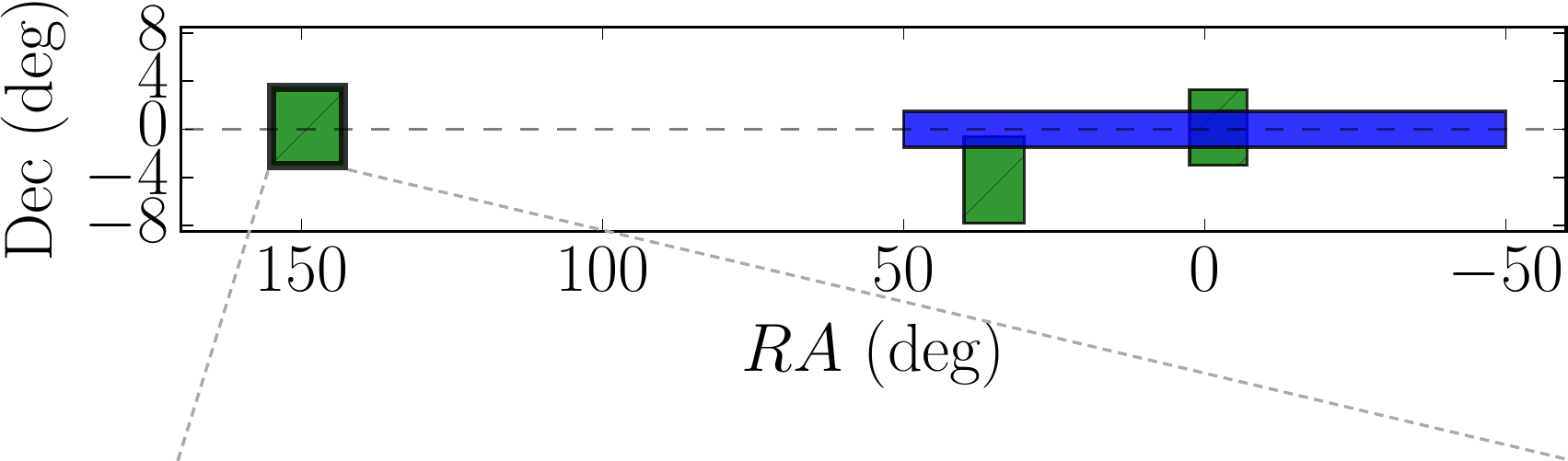}

\vspace{-0.5mm}

\includegraphics[width=80mm]{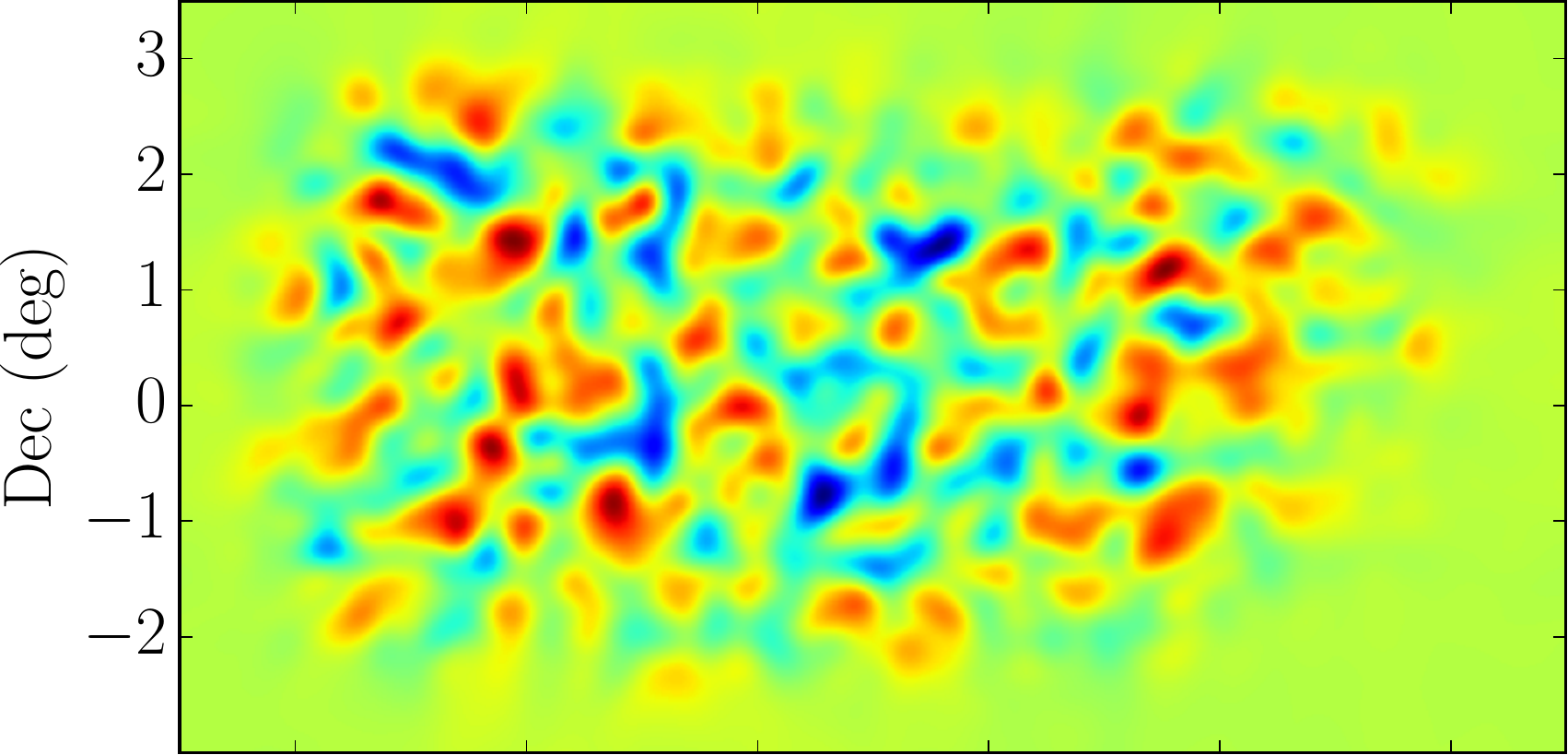}

\vspace{-0.5mm}

\includegraphics[width=80mm]{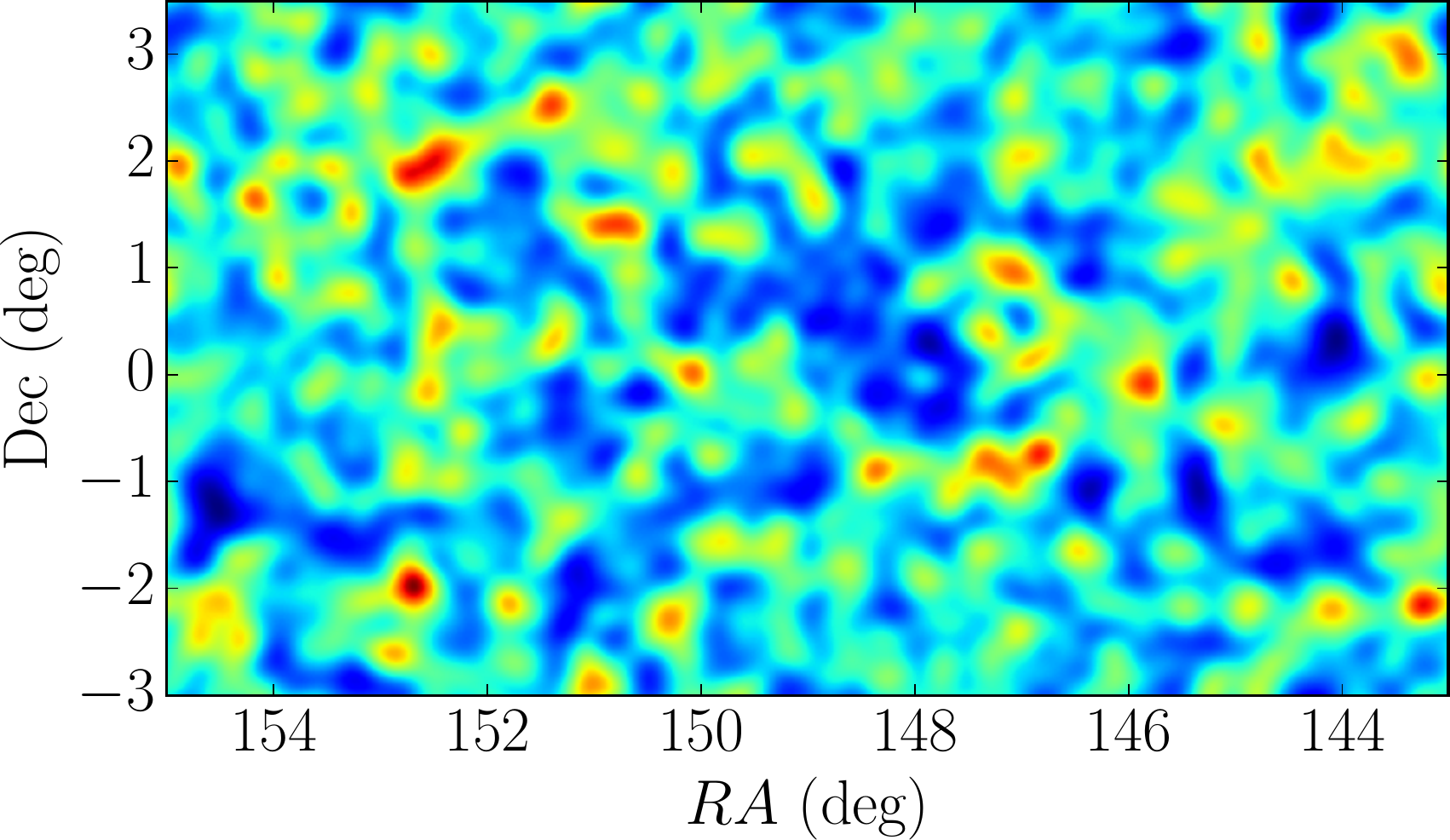}
\caption{{\it Top panel:} Footprint of the patches used in this analysis: ACT ({\it blue}) and ACTPol ({\it green}, left to right: D1, D6, D5). {\it Middle panel:} ACTPol D1 lensing convergence map $\kappa$ smoothed to suppress power below $20'$ scales. The spatial modulation, primarily due to the windowing of the temperature and polarization maps by the pixel weight map, is evident. The ACTPol lensing convergence is noise dominated for scales $\lesssim 1$~degree. {\it Bottom panel:} The FIRST overdensity field $g$ over the same patch, smoothed to the same scale, is noise dominated at all scales.}
\label{figMap}
\end{figure}

We also fold in the three ACTPol `deep' fields from the first-season dataset, labelled D1, D5 and D6, with a temperature white-noise component of $16.2, 13.2, 11.2~\mu$K-arcmin, respectively, over a total area of 206 deg$^2$ (37 deg$^2$ of which overlaps with the ACT Equatorial strip). All maps in this analysis use $0.5' \times 0.5'$ pixels, and their positions are shown in Fig.~\ref{figMap}.

For each ACTPol patch we use the minimum-variance (MV) linear combination of the reconstructed convergence maps estimated from each quadratic pair (TT, TE, EE, EB). Following the \cite{Polarbear:2013}, \cite{vanEngelen:2014} and \cite{Story:2014} we perform the combination in Fourier space, weighting each convergence map by the mode-dependent inverse-variance noise to obtain the MV combination. All lensing convergence maps are mean-field subtracted to remove the lensing-like effect at large scales of mode-coupling from the windowing of the temperature and polarization fields.

As described in \cite{vanEngelen:2014} an apodization window is applied to the ACT and ACTPol temperature and polarization map prior to lensing reconstruction. This windowing operation includes a cosine taper at the map edges to remove discontinuous edges and weighting by the pixel hitmap to optimize the signal-to-noise of the reconstruction. The resulting quadratic estimator reconstruction is therefore also windowed, resulting in a scale-dependent suppression of power. Following \cite{Bleem:2012}, \cite{Sherwin:2012}, \cite{Hand:2013} and \cite{vanEngelen:2014}, we use realistic Monte-Carlo simulations to calculate the transfer function correction by computing the mean cross-spectrum between noiseless lensing realizations and their corresponding reconstructions within the lensing pipeline. This correction ($<5\%$ for ACTPol, $\approx 10\%$ for ACT) is then applied to the maps when computing the data cross-spectrum to account for the suppression of power due to windowing. 

\subsection{FIRST}
\label{ssFIRST}

The FIRST survey \citep{Becker:1995} was carried out between 1993 and 2011 at 1.4 GHz with the VLA in B configuration. The final catalogue \citep{Helfand:2015} contains 946,432 sources covering 10,575 deg$^2$, with an angular resolution of $5.4''$ (FWHM) and to a completeness of $95\%$ at flux $S_{1.4 {\rm GHz}} > 2$ mJy. The oblique decision-tree program developed by the FIRST survey team \citep{White:1997} determines the probability that each catalogue entry is the result of a spurious sidelobe response to a nearby bright source. We exclude entries with a sidelobe probability of $>0.1$, leaving 720,219 sources above 1 mJy.

To address the issue of extended radio sources resulting in multiple detections for one host galaxy, perhaps none of which corresponds to the core itself (and therefore any associated optical source), we have followed \cite{Cress:1996} in applying a collapsing radius of $72''$ (0.02 deg) to the FIRST catalogue. Any FIRST sources within this radius of one another are grouped and combined to form a single entry, positioned at the flux-weighted average coordinates of the group, and attributed with their total flux density. Around $32\%$ of all FIRST sources are collapsed (in groups of average size 2.3 sources per group), forming $17\%$ of the resulting catalogue. These multiple-component sources will come from AGN, which dominate the source population at high flux density and high redshift; at lower flux density and redshift, starbursts and normal star-forming galaxies (SFGs) are increasingly dominant \citep{Condon:2002}. 

The redshifts of individual sources are not determined by FIRST, but can be found by identifying counterparts in SDSS, which gives redshifts for the brighter, nearby sources in the FIRST sample. The closest sources are most likely star-forming galaxies  \citep[e.g.,][]{Condon:2002, Wilman:2008}; by removing them we simplify the measurement as a constraint on the bias of the dominant astrophysical population (i.e. AGN). 

We do this by initially taking all sources in the catalogue which lie in the ACT and ACTPol patches described above ($\approx$ 38,000 sources). AGN dominate the radio luminosity function for $L_{1.4 {\rm GHz}} > 10^{23}$ W Hz$^{-1}$ \citep{Condon:2002, Jarvis:2004, Mauch:2007}. Given the flux limit of FIRST (1mJy), and assuming a spectral index of $\alpha = 0.8$ ($S_\nu \propto \nu^{-\alpha}$) for the AGN, this luminosity threshold corresponds to sources above redshift $z = 0.2$. We identify optical matches to the radio sources within SDSS, treating as reliable all matches within $2''$ of the radio source following \cite{Lindsay:2014a}. Given the density of SDSS sources, the level of spurious optical matches identified with this technique is below $2\%$.  A fraction of $0.27$ of the FIRST sources in the lensing fields have an optical match obtained in this manner. We remove all sources with a known redshift below $z = 0.2$, constituting $18\%$ of the sources with a reliable redshift, or $5\%$ of the total number of sources. Given the small fraction of sources removed, this procedure has only a small effect on the results (Section~\ref{sResults}). The final sample comprises $\approx$ 36,000 sources with a mean angular density of $71$ sources deg$^{-2}$.

Within each ACT and ACTPol patch a corresponding map of the overdensity of sources $g$ is produced in a similar way to \cite{Sherwin:2012} and \cite{Geach:2013}. We create a map at the same pixelation as the lensing map and define the radio-galaxy overdensity map $g$ by
\begin{equation}
g_i = \frac{n_i}{\bar{n}} - 1,
\end{equation}
where $n_i$ is the number of sources in each pixel and $\bar{n}$ is the mean number of galaxies per pixel. This overdensity map is then smoothed with a Gaussian with FWHM of $2'$ to obtain a well-defined pixel window function. 

\subsection{Analysis methods}
\label{ssAnalysis}
We compute the cross-spectrum between the lensing convergence from ACT and ACTPol with the FIRST radio galaxy overdensity.

Following the procedures outlined in \cite{Das:2011} and \cite{Hand:2013}, we correct for mode-coupling induced by windowing in real space and from applying annular binning in Fourier space, computing an unbiased estimator of the binned cross-spectrum $C_b^{\kappa g}$. The binning we adopt is given in Table~\ref{tabBins}.

To determine the full band-power covariance matrix we cross-correlate realistic simulations of the reconstructed lensing fields with the radio source maps (which are in principle uncorrelated). Production of these realistic simulations is described in \cite{Das:2011} and \cite{vanEngelen:2014}.

\begin{figure}
\centering
\includegraphics[width=85mm]{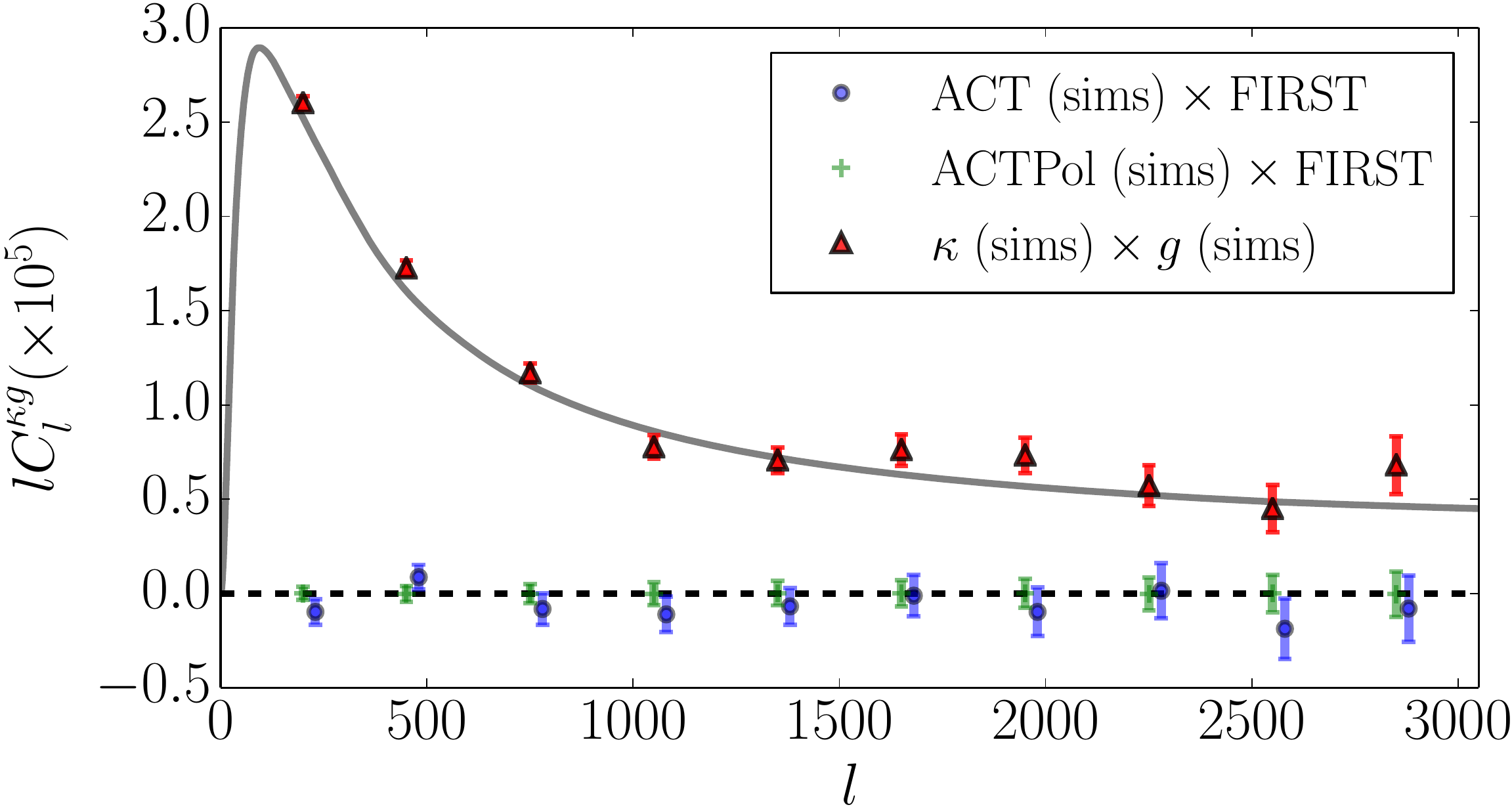}
\caption{{\it Pipeline validation:} Mean cross-spectrum $l C_l^{\kappa g}$ of the FIRST radio source map with ACTPol lensing simulations ({\it green dashes}, $N_{\rm sims}=2048$) and ACT lensing simulations ({\it blue circles},  $N_{\rm sims}=480$) as described in \ref{ssAnalysis}. We displace the ACT points by $\Delta l = 30$ to the right for visual clarity. The measurements are consistent with null, demonstrating that our pipeline does not induce spurious cross-power in the absence of correlation. Error bars shown are the diagonal components of the empirical covariance matrix derived from the same Monte Carlo simulations, scaled appropriately by $\sqrt{N_{\rm sims}}$. We also show the recovered mean cross-spectrum from realistic {\it correlated} simulations ({\it red triangles}, Section \ref{ssAnalysis}). We cross-correlate input convergence maps, which have added scale-dependent Gaussian noise, with correlated realizations of a galaxy field. This demonstrates that our pipeline is able to recover in an unbiased fashion a known input cross-spectrum (although we note this does not test the lensing reconstruction pipeline, for which we refer to the systematic tests in \protect\cite{vanEngelen:2014}). The generative model for the cross-spectrum is not the fiducial cross-spectrum, but this is unimportant for the purposes of this test.}
\label{figNull}
\end{figure}

This procedure ignores the cosmic variance contribution to the uncertainties in the data coming from the correlated part of the two maps, $C_l^{\kappa g}$. We neglect this as both maps are noise-dominated at the relevant scales for this analysis. Bin-to-bin correlations are $<10\%$ for all off-diagonal elements of the covariance matrix. We also check that the mean cross-spectrum is consistent with null (Figure~\ref{figNull}), confirming that our pipeline does not induce spurious cross-power in the absence of correlation. 

Approximately $50\%$ of the ACTPol D5 patch and $15\%$ of D6 overlap with the ACT Equatorial strip (Fig.~\ref{figMap}). There is therefore a correlation between the ACT and ACTPol cross-spectra, as common CMB modes in the primary temperature map have been used to reconstruct the lensing convergence over these regions. Noiseless temperature maps from ACT and ACTPol, and negligible polarization information from ACTPol, would result in a perfect correlation between the reconstructed convergence maps. However, this overlapping area represents $37$ deg$^2$ of the total $470$ deg$^2$ of this analysis, and hence at most a 4\% overestimate of the detection significance, which we neglect given the statistical errors. We thus average the ACT and ACTPol data cross-spectra with inverse-variance weighting.

In order to check for bias in the cross-spectrum estimator, we ran 500 pairs of simple simulated convergence and radio density maps through the cross-correlation pipeline, generating new {\it correlated} simulations. To obtain these pairs we draw as signal maps aperiodic correlated Gaussian realizations from power spectra obtained assuming {\sl Planck} best-fit cosmological parameters and a fiducial bias model and source distribution for the radio galaxies \citep{Kamionkowski:1997, PlanckXVI}. We add Gaussian noise realizations to the convergence maps, appropriate for the temperature sensitivity of ACT (Section \ref{ssACT}), using the formalism of \cite{Hu:2002} to calculate the reconstruction noise. ACTPol maps are less noisy, but the precise noise level is unimportant for this test. For each pixel $i$ in the radio signal map $g$ we draw a Poisson random variable $X_i$ with mean $\bar{n}(1 + g_i)$, where $\bar{n}$ is the average number of sources per pixel. We set $\bar{n} = 71 $~sources deg$^{-2}$ to reflect the source density in the data. We then redefine $g_i \leftarrow X_i/\bar{n} - 1$ and finally smooth the resulting map with a Gaussian beam of FWHM $2'$. 

These simulated maps, by construction, have signal, noise and correlation properties which mimic the data, although they do not have the full spatially anisotropic noise properties. These lensing simulations have not been processed through the lensing reconstruction pipeline, but here we use them simply for checking bias in the cross-correlation pipeline. We refer to the systematic tests in \cite{vanEngelen:2014} for checks of the lensing pipeline. We find that we do not require apodization of the maps to produce the observed unbiased results; the mean auto- and cross-spectra of these simulations are consistent with the assumed input spectra (Fig.~\ref{figNull}). 

\subsection{Modelling}
\begin{figure}
\centering
{\includegraphics[width = 80mm]{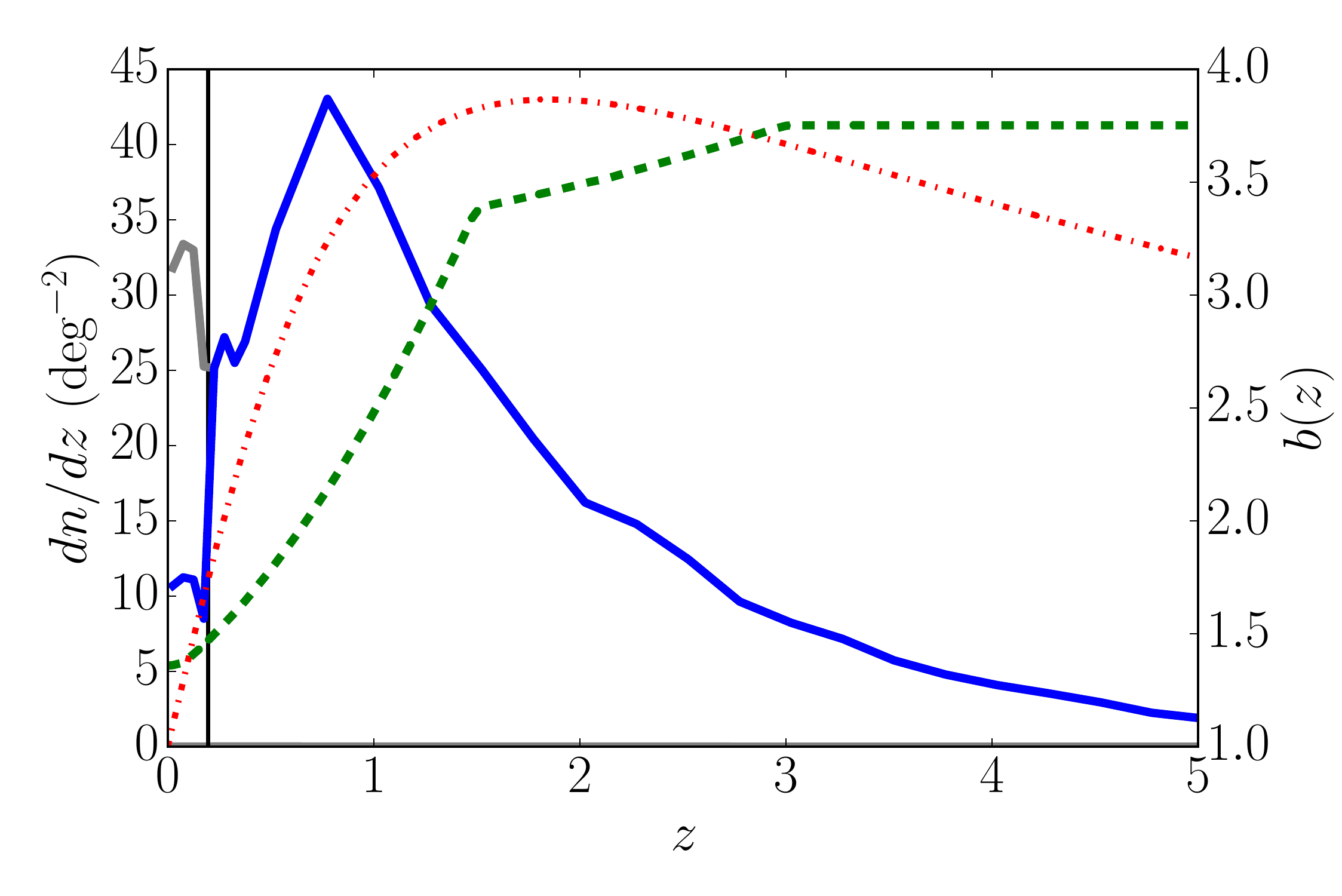}}
\caption{{\it Green dashed:} Fiducial bias model $b(z)$, constructed as a weighted average of the constituent source population bias models (Section~\ref{ssFIRST}). {\it Blue solid:} Source redshift distribution model as derived from SKADS and including the cut of a fraction of the $z<0.2$ sources.{\it Vertical black line:} $z = 0.2$. {\it Grey solid:} As previous but without the redshift cut. {\it Red dot-dashed:} Unnormalised CMB lensing kernel $W_\kappa(z)$.}
\label{figModels}
\end{figure}
The theoretical cross-spectrum can be written under the Limber approximation as
\begin{equation}
C_l^{\kappa g} = \int_0^\infty dz \frac{H(z)}{\chi^2(z)}W_\kappa(z) W_g(z) P\left(\frac{l}{\chi(z)}, z \right),
\label{eqPS}
\end{equation}
where $H(z)$ is the Hubble parameter, $\chi(z)$ is the comoving distance to redshift $z$, $P(k, z)$ is the non-linear matter power spectrum (wavenumber $k = l/\chi$) and $\{W_i\}$ are the appropriate kernels for the two dark matter probes $\kappa$, $g$. The dominant term in $W_g$ is directly proportional to the tracer bias $b(z)$:
\begin{equation}
W_g(z) = b(z)\frac{dn}{dz} + M(z),
\label{eqKernel}
\end{equation}
where $dn/dz$ is the normalised source redshift distribution and $M(z)$ is a sub-dominant contribution from the magnification bias (see e.g., \cite{Sherwin:2012} for the full expression). The magnification bias term is independent of the tracer bias, and for the FIRST sources in our sample is small ($< 6\%$ of the total). A rescaling of the bias amplitude therefore corresponds linearly to a rescaling of the cross-spectrum $C_l^{\kappa g}$. We compute the theory $P(k,z)$ using best-fit {\sl Planck} cosmological parameters, including non-linear corrections using CAMB with Halofit \citep{Lewis:1999, Smith:2003,Takahashi:2012}. 

We use the SKA Design Study (SKADS) simulated radio continuum catalogue to construct a fiducial bias model $b(z)$ and redshift distribution $dn/dz$ for the $S_{1.4{\rm GHz}} > 1$~mJy radio sources \citep[see][for details]{Wilman:2008, Wilman:2010}. The simulation lacks the mass resolution to directly resolve galaxy- and group-sized haloes for a robust implementation of the galaxy clustering, but the source counts, redshift distribution, and variations in space density are defined by extrapolating observed luminosity functions, and implementing a bias model, for each of five individual radio populations: AGN (FRI and FRII types, radio-quiet quasars), normal star-forming galaxies and starburst galaxies. These populations are assigned a single halo mass each, used to define $b(z)$ as described by \cite{Mo:1996}, with the bias held fixed above a particular redshift to prevent unphysical clustering where the bias is poorly constrained observationally \citep[see Figure 3 of][]{Raccanelli:2012}. The simulated catalogue informs us how the relative numbers of these populations evolve with redshift, and how the observed bias will evolve accordingly for a mixed sample of sources.

By comparing the distribution of known source redshifts (Section~\ref{ssFIRST}) with the SKADS simulation, we find an estimated $66\%$ of low-redshift sources are removed by the $z<0.2$ cut. To construct $dn/dz$ we therefore weight $z<0.2$ sources by $0.34$ relative to higher-redshift sources, similar to the approach of \cite{Lindsay:2014b}. After the redshift cut $\approx 96\%$ of the sources in our sample are expected to be AGN, with a $\approx$ 4\% contamination fraction of star-forming and starburst galaxies, and we estimate the final sample to have a median redshift $\tilde{z} = 1.3$. These fiducial models are shown in Figure~\ref{figModels}. We discuss the limitations of this model, including the effect of not removing the low-redshift sources, in Section~\ref{ssLimitations}. 

Finally, we bin the theoretical cross-spectrum as for the data, accounting for the mode-coupling matrix of each patch \citep{Das:2011}. We then compare the model to the data using a simple Gaussian likelihood, and primarily fit for an overall scaling $A$ to the fiducial bias model, such that $b(z) \rightarrow Ab(z)$.
\section{Results}
\label{sResults}
\begin{table}
\begin{adjustwidth}{-0.4cm}{}
\begin{center}
\begin{tabular}{ccrrr}
\hline \hline
Bin $b$ & $[l_{\rm min},$ & $C_{b,{\rm ACT}}^{\kappa g}$ & $C_{b,{\rm ACTPol}}^{\kappa g}$ & $C_{b,{\rm comb}}^{\kappa g}$    \\ 
 &  $l_{\rm max}]$& $(\times 10^7)$ & $(\times 10^7)$ & $(\times 10^7)$  \\ 
\hline
200 & [$100$,$300$] & $1.76 \pm 0.74$ & $2.04 \pm 0.80$ & $1.89 \pm 0.54$ \\
450 & [$301$,$600$] & $0.59 \pm 0.32$ & $-0.20 \pm 0.42$ & $0.30 \pm 0.25$ \\
750 & [$601$,$900$] & $0.57 \pm 0.25$ & $0.32 \pm 0.31$ & $0.48 \pm 0.19$ \\
1050 & [$901$,$1200$] & $0.28 \pm 0.20$ & $0.38 \pm 0.26$ & $0.32 \pm 0.16$ \\
1350 & [$1201$,$1500$] & $0.17 \pm 0.16$ & $-0.08 \pm 0.22$ & $0.09 \pm 0.13$ \\
1650 & [$1501$,$1800$] & $0.04 \pm 0.15$ & $0.01 \pm 0.19$ & $0.03 \pm 0.12$ \\
1950 & [$1801$,$2100$] & $0.11 \pm 0.14$ & $-0.00 \pm 0.18$ & $0.07 \pm 0.11$ \\
2250 & [$2101$,$2400$] & $0.07 \pm 0.14$ & $-0.01 \pm 0.17$ & $0.04 \pm 0.11$ \\
2550 & [$2401$,$2700$] & $0.06 \pm 0.14$ & $0.19 \pm 0.18$ & $0.11 \pm 0.11$ \\
2850 & [$2701$,$2999$] & $-0.01 \pm 0.13$ & $-0.23 \pm 0.19$ & $-0.08 \pm 0.11$ \\
\hline
\end{tabular}
\end{center}
\end{adjustwidth}
\caption{The measured cross-spectrum $C_{b,{\rm ACT}}^{\kappa g}$ for FIRST radio sources with ACT and ACTPol lensing. The bins are chosen to be wide enough that correlations are small ($<10\%$, Section~\ref{ssAnalysis}), but narrow enough to resolve structure in the cross-spectrum. }
\label{tabBins}
\end{table}
\begin{figure}
\centering
\includegraphics[width=85mm]{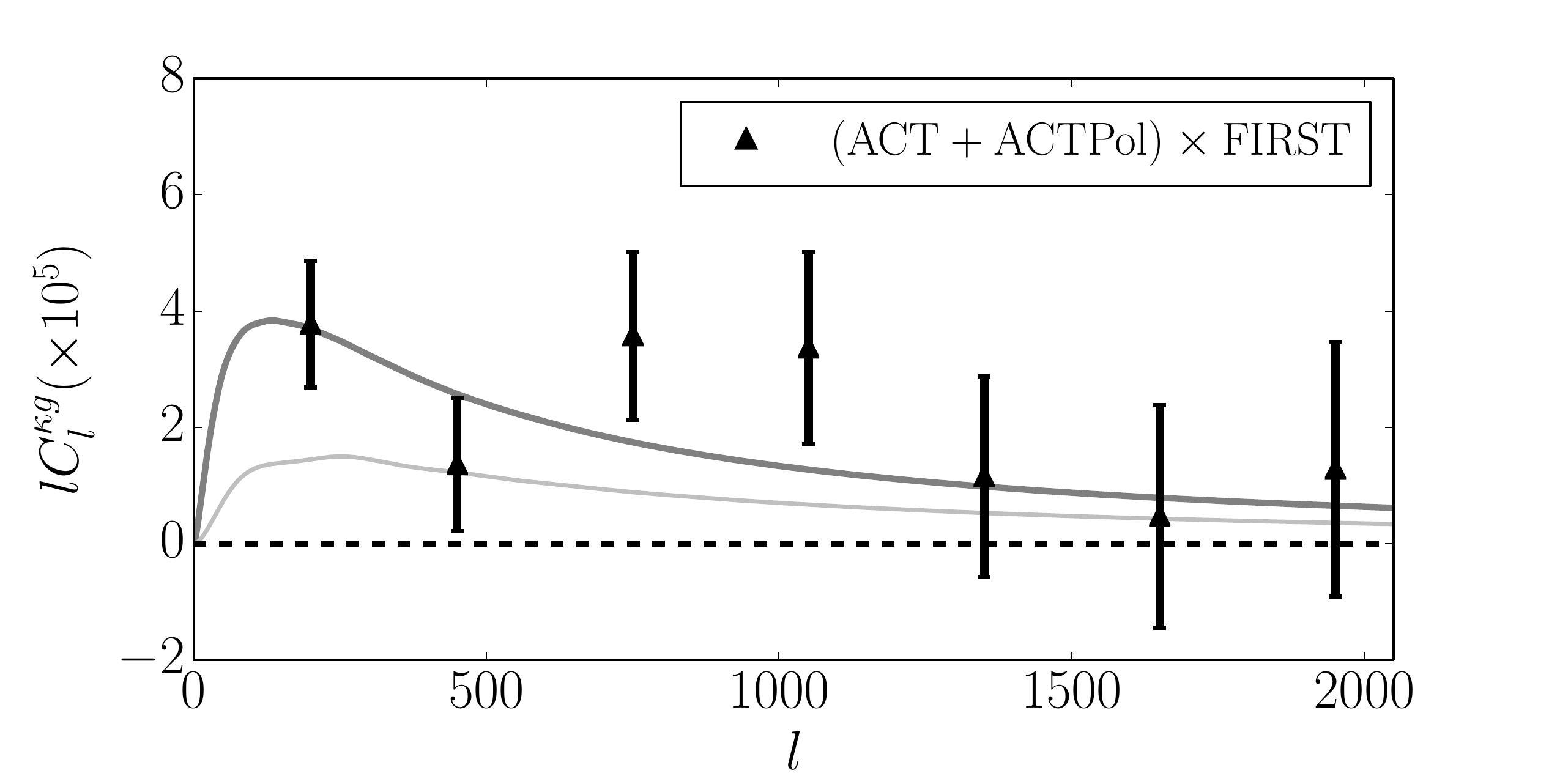}
\caption{Data cross-spectrum $l C_l^{\kappa g}$ for (ACT+ACTPol) $\times$ FIRST. {\it Dark grey, solid}: the best-fit cross-spectrum. {\it Light grey, solid}: the contribution from $z>1.5$ sources. We restrict the plot to $l<2000$ where the signal-to-noise dominates. We show as error bars the diagonal components of the empirical covariance matrix derived from Monte Carlo simulations (Section \ref{ssAnalysis}). Scaling the amplitude of the fiducial bias model, the combined significance of the bias detection is $4.5 \sigma$ (Section \ref{sResults}).}
\label{figCOMBdata}
\end{figure}
\begin{figure}
\centering
\includegraphics[width=85mm]{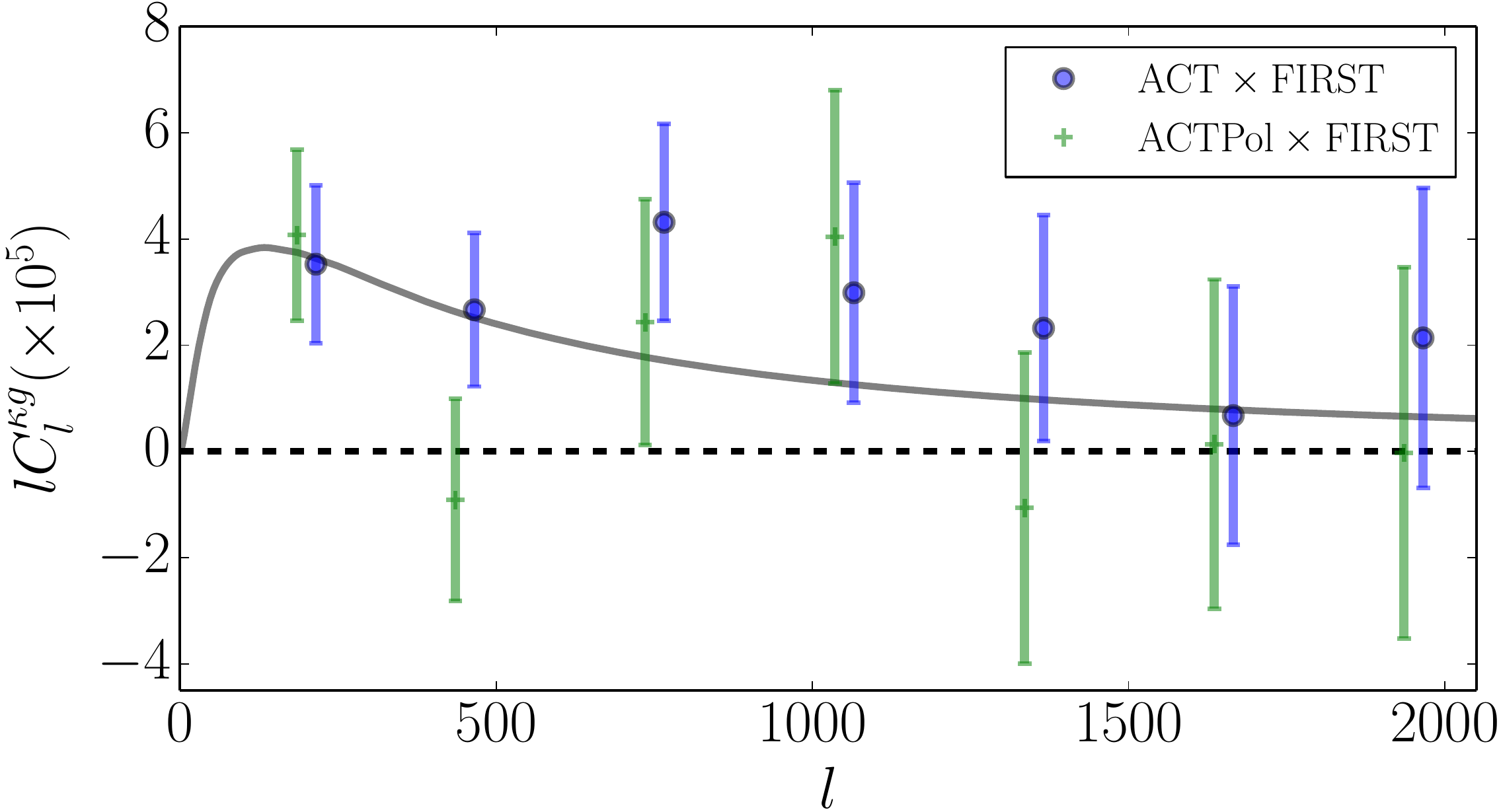}
\caption{Data cross-spectrum $l C_l^{\kappa g}$ for ACT $\times$ FIRST ({\it blue}) and ACTPol $\times$ FIRST ({\it green}). {\it Dark grey, solid}: the best-fit cross-spectrum for the combined data. {\it Light grey, solid}: the contribution from $z>1.5$ sources. The ACT and ACTPol points have been displaced to the right and left by $\Delta l = 15$, respectively, for visual clarity. We show as error bars the diagonal components of the empirical covariance matrix derived from Monte Carlo simulations (Section \ref{ssAnalysis}).}
\label{figACTFIRSTdata}
\end{figure}
The cross-spectra for ACT $\times$ FIRST, ACTPol $\times$ FIRST, and their combination are shown in Figs. ~\ref{figCOMBdata}~and~\ref{figACTFIRSTdata} and reported in Table~\ref{tabBins}. We find $A_{\rm ACT} = 1.22 \pm 0.31$ and $A_{\rm ACTPol} = 0.85 \pm 0.36$, with combined constraint $A = 1.06 \pm 0.24$. The goodness-of-fit statistics for these best-fit models are reported in Table 2. This amplitude is consistent with the expected bias from the radio simulations; we interpret the result further in Section \ref{sBias}. 

\begin{table}
\begin{center}
\begin{tabular}{|c|c|c|c|c|}
\hline \hline
 & $A$ & $S/N$ & $\chi^2$ \hspace{0.1mm} $(\nu)$ & PTE  \\ 
\hline
ACT & $1.22 \pm 0.31$ & 3.9 & 3.2 (9) & 0.96 \\ 
ACTPol & $0.85 \pm 0.36$ & 2.4 & 7.2 (9) & 0.62 \\ 
\hline
Comb. & $1.06 \pm 0.24$ & 4.5 & 11.0 (19) & 0.92 \\ 
\hline
\end{tabular}
\caption{Results showing the bias amplitude $A$ relative to the fiducial model of Fig.~\ref{figModels}. We also quote the signal-to-noise ratio $S/N$, Chi-squared values at the best-fit $\chi^2$, the number of degrees of freedom $\nu$ and the probability to exceed this $\chi^2$ (PTE) under the assumption of the best-fitting model.}
\end{center}
\label{tabPTEs}
\end{table} 
The parameter $A$ only scales the bias-dependent part of the theoretical model. To assess the overall detection significance we rescale the amplitude of the total theoretical cross-spectrum by a free parameter $\alpha$: $C_l^{\kappa g} \rightarrow \alpha C_l^{\kappa g}$. This is equivalent to equally rescaling both terms in Eq.~\ref{eqKernel}, including the magnification bias term. The combined data require $\alpha = 1.06 \pm 0.24$, and the cross-spectrum is detected at $\sqrt{\chi^2_{\rm null} - \chi^2_{\rm bf}} = 4.4\sigma$ statistical significance. Here $\chi^2_{\rm null}$ = 31.4 is the chi-squared value of the fit under the null hypothesis (no cross-correlation) and $\chi^2_{\rm bf}$ = 11.0 is the chi-squared value for the best-fit model (number of degrees of freedom $\nu = 19$). 

The mean cross-spectrum of ACT and ACTPol Monte-Carlo simulations with the FIRST dataset is shown in Fig.~\ref{figNull} and is consistent with null (Section~\ref{ssAnalysis}). These simulations reproduce the amplitude and statistics of the lensing field but not the true mass distribution on the sky. 

We further test our pipeline, checking for spurious correlations present only in the lensing and galaxy data, by performing two additional null tests.

\begin{figure}
\centering
\includegraphics[width=85mm]{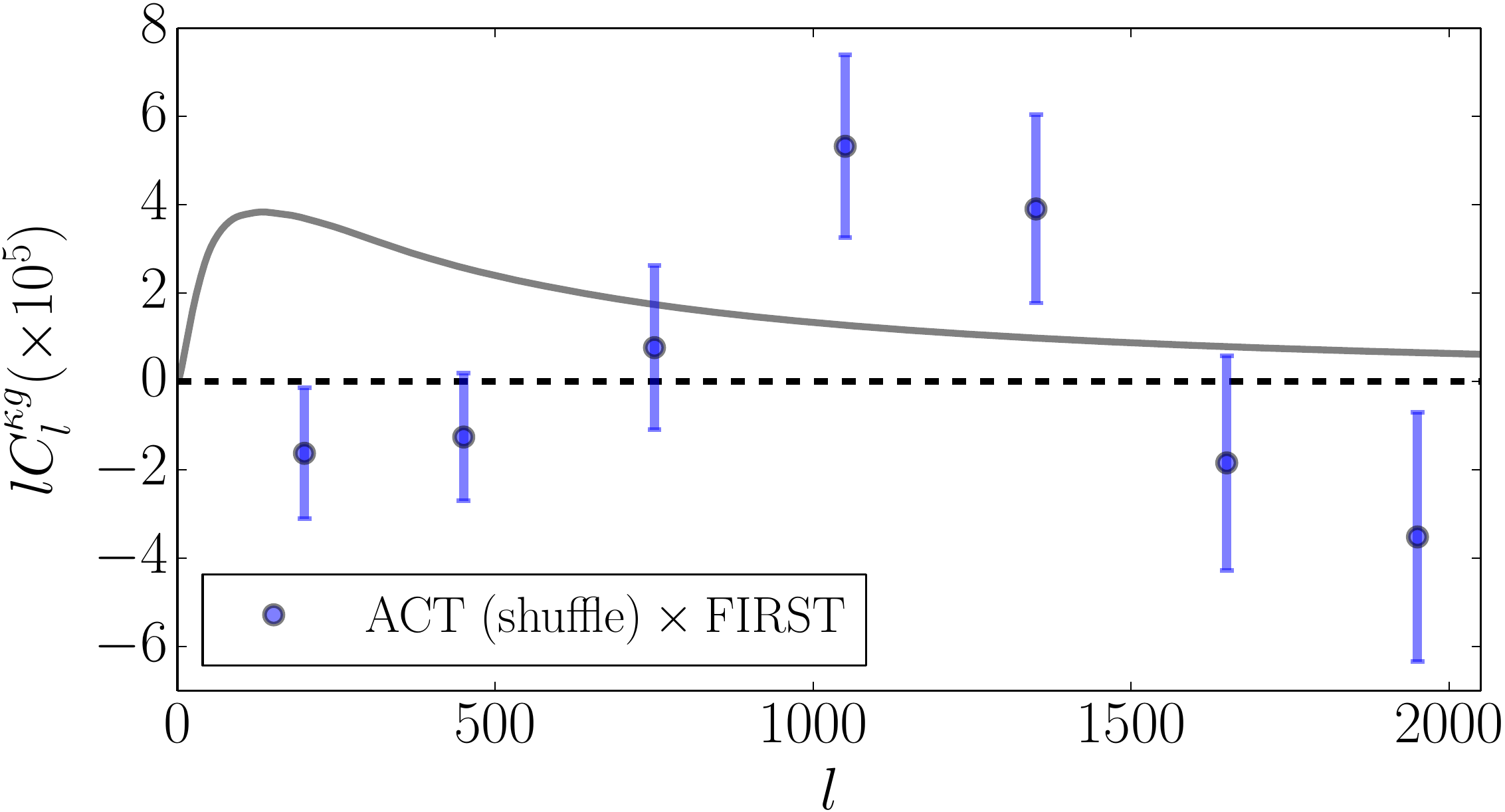}
\caption{Cross-correlation between shuffled FIRST maps with ACT lensing convergence. Fitting the normalization of the fiducial bias model $A$ to these data, we obtain $A_{\rm ACT, shuffle} = -0.18 \pm 0.31$, consistent with null (Section~\ref{sResults}). {\it Grey solid curve:} Cross-spectrum for the fiducial bias model which best fits the data of Fig.~\ref{figCOMBdata}.}
\label{figShuffle}
\end{figure}

First, we randomly permute the six FIRST patches within the equatorial strip, such that all patches are moved from their true position, with respect to the fixed ACT patches. We recompute the cross-spectrum, shown in Fig.~\ref{figShuffle}. Fitting the normalization of the fiducial bias model $A$ to these data, we obtain $A_{\rm ACT, shuffle} = -0.18 \pm 0.31$, consistent with null. The chi-squared value of the null hypothesis is $18.2$ for $\nu = 10$ degrees of freedom, or a probability-to-exceed the observed chi-squared of $5\%$. 

\begin{figure}
\centering
\includegraphics[width=85mm]{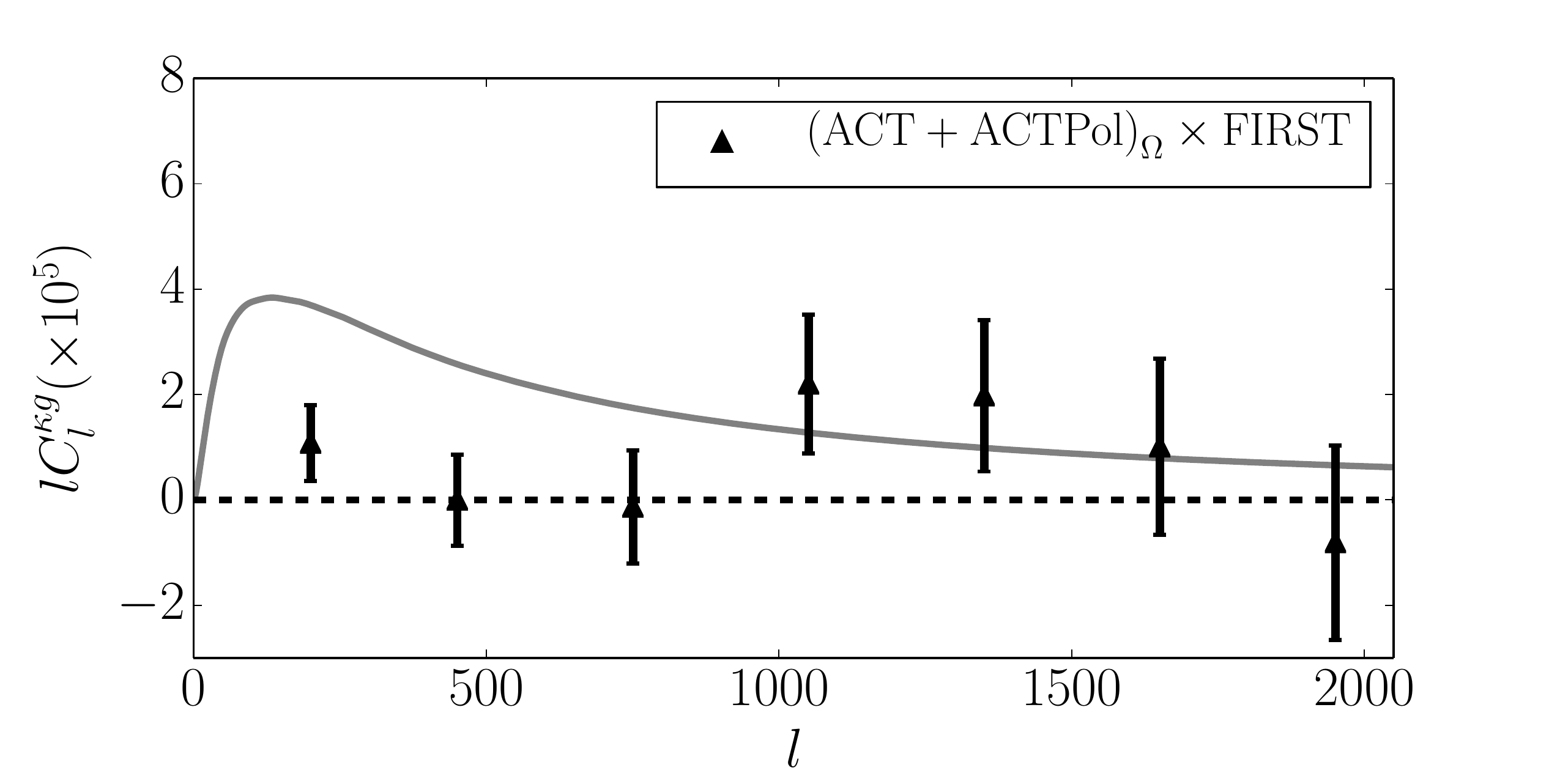}
\caption{Cross-spectrum $l C_l^{\kappa g}$ between ACT+ACTPol lensing curl maps $\Omega$ and FIRST (Section~\ref{sResults}). We restrict to $l<2000$ for comparison with Fig.~\ref{figCOMBdata}. Fitting the normalization of the fiducial bias model $A$ to these data, we obtain $A_{\Omega} = 0.19\pm 0.17$ and a chi-squared value of the null hypothesis of $21.8$ for $\nu = 19$ degrees of freedom (a probability-to-exceed of $0.29$). As expected this cross-correlation is consistent with null. {\it Grey solid curve:} Cross-spectrum for the fiducial bias model which best fits the data of Fig.~\ref{figCOMBdata}.}
\label{figCurlNull}
\end{figure}

Second, we make reconstructions of the lensing field where the deflection field has been redefined - as the curl of the lensing potential - and hence the expected `convergence' $\Omega$ is zero, following \cite{Sherwin:2012} and \cite{vanEngelen:2014}. These maps contain reconstruction noise but should contain no common signal with the overlapping galaxy field. We recompute the cross-spectrum of the lensing curl maps $\Omega$ with the FIRST maps, shown in Fig.~\ref{figCurlNull}. Fitting the normalization of the fiducial bias model $A$ to these data, we obtain $A_{\Omega} = 0.19 \pm 0.17$. Error bars are calculated from the data auto-spectra using the Knox formula \citep{Knox:1995}. The chi-squared value of the null hypothesis is $21.8$ for $\nu = 19$ degrees of freedom, or a probability-to-exceed of $0.29$, confirming a null result. 

Removal of the known $z<0.2$ sources, which constitute $\approx 5\%$ of the FIRST sample (Section~\ref{ssFIRST}), has only a small effect on the inferred bias amplitude: without removal we find a combined constraint $A_{\rm noZcut} = 1.08 \pm 0.24$, consistent with expectations given the shape of the lensing kernel and low bias of SFGs at low redshift. 
\begin{figure}
\centering
\includegraphics[width=82mm]{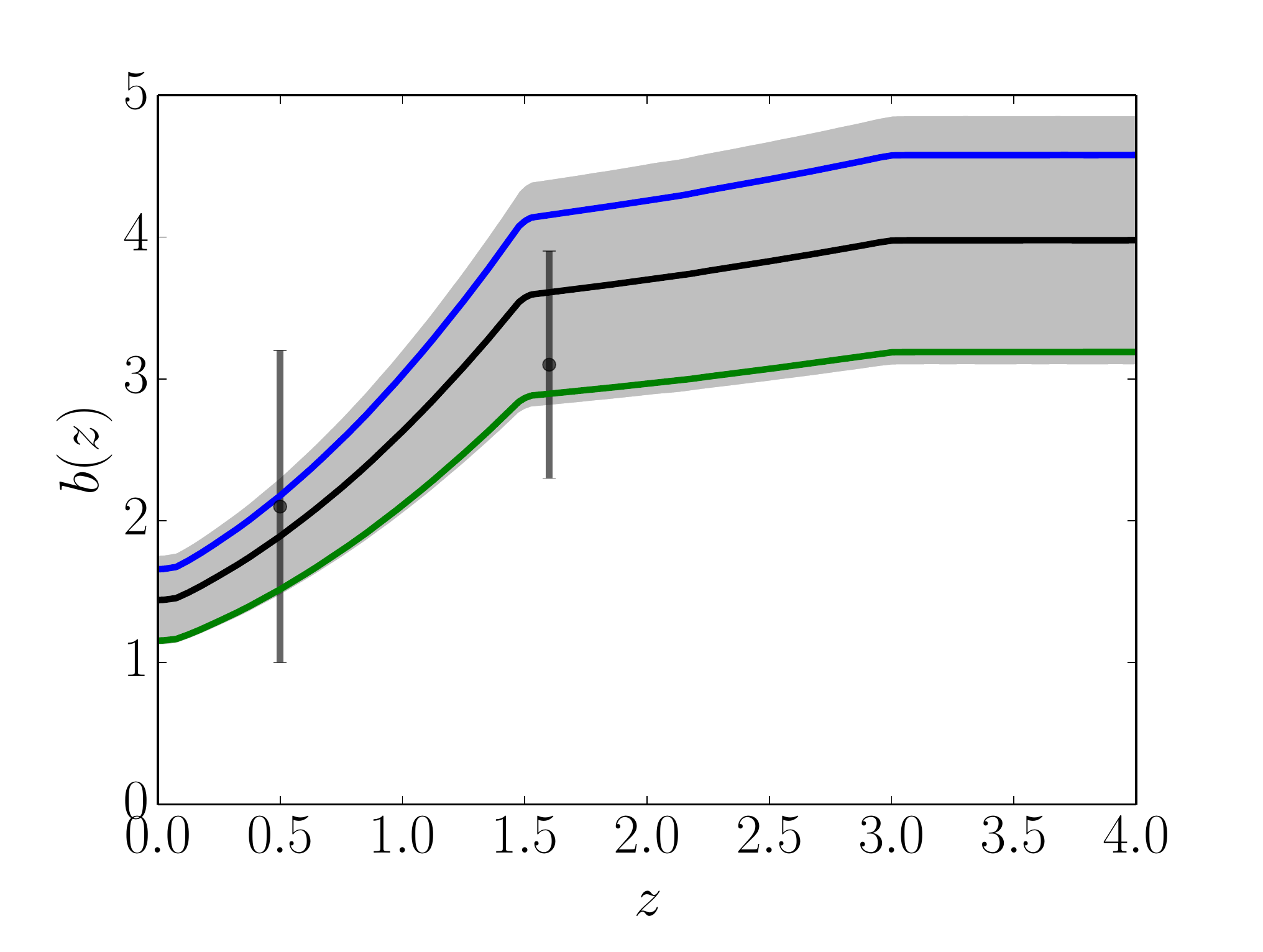}
\caption{{\it Solid lines}: Fiducial bias model scaled by the best-fit bias amplitude for ACT $\times$ FIRST ({\it blue}), ACTPol $\times$ FIRST ({\it green}) and for the combined result ({\it black}). The grey shaded area shows the corresponding $1\sigma$ credible region for the combined result (Table~\ref{tabPTEs}). Note we constrain only the {\it normalization} of the fiducial bias model; the grey band is indicative of this uncertainty only. Using the combined measurements the bias at the effective redshift of the analysis is $b(z_{\rm eff} = 1.5)= 3.5 \pm 0.8$ (Section~\ref{sResults}). {\it Data points:} Bias inferred from sources with known individual redshifts ({\it left}) and with unknown individual redshifts ({\it right}), plotted at the effective redshift of each measurement (Section~\ref{sBias}).}
\label{figPosterior}
\end{figure}
\subsection{AGN Bias}
\label{sBias}
\begin{figure}
\centering
{\includegraphics[width = 70mm]{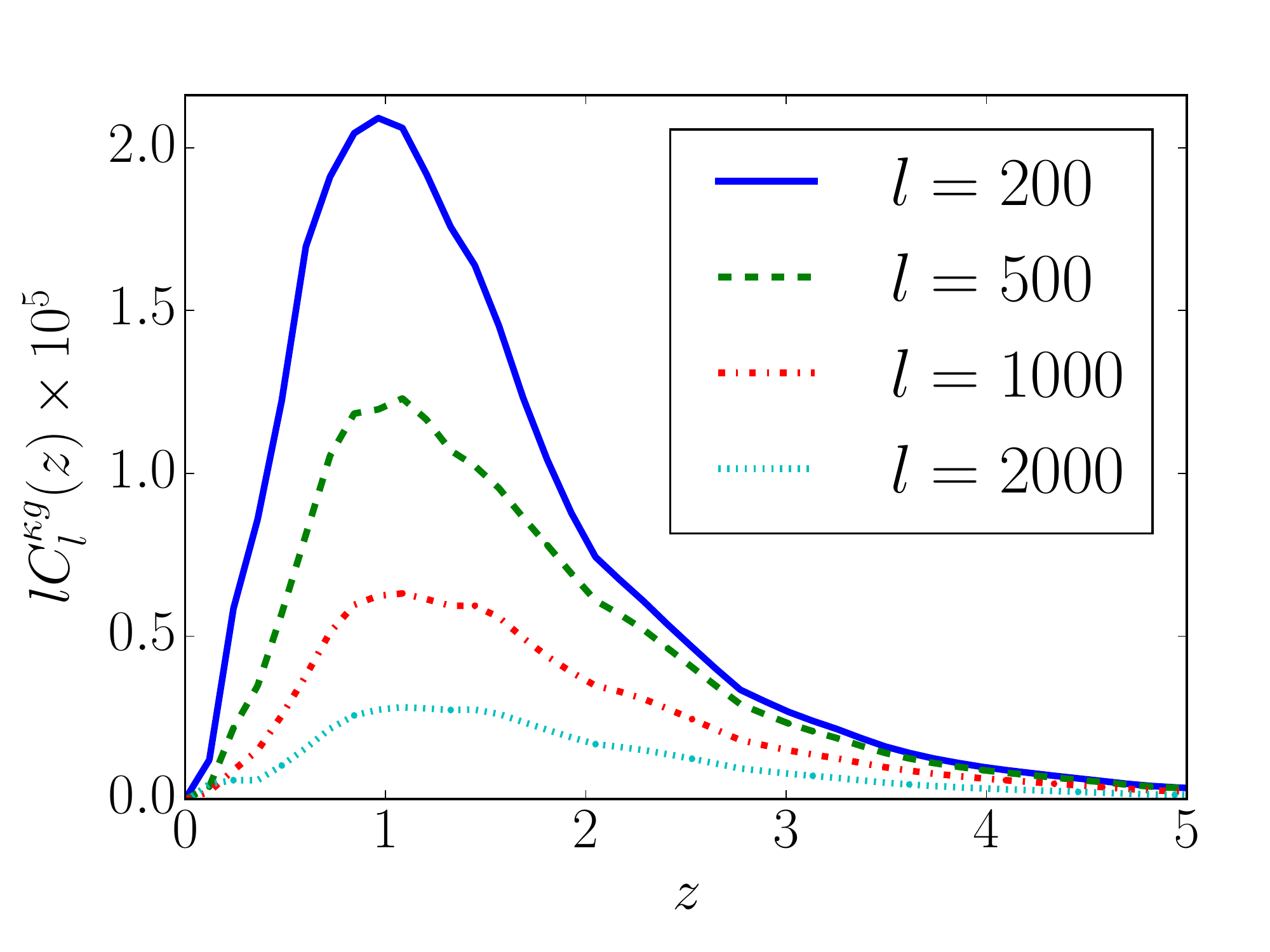}}
\caption{Cross-power spectrum kernel $C_l^{\kappa g}(z) \equiv W_\kappa(z) W_g(z) P(l/\chi(z);z)$, demonstrating the scale-dependent sensitivity of the cross-spectrum to source redshift. At $l = 200$, where the signal-to-noise peaks, the mean redshift of the kernel is $\int z C_{200}^{\kappa g}(z) dz / C_{200}^{\kappa g}= 1.5$, which we adopt as the effective redshift $z_{\rm eff}$ of the radio source bias measurement. The spread in the kernel reflects sensitivity to a wide range of redshift. See Section~\ref{ssAnalysis} for details.}
\label{figKernels}
\end{figure}
The fiducial bias model scaled by our best-fit amplitude is shown in Figure \ref{figPosterior}. We determine the redshifts to which our measurement is most sensitive by considering the kernel $C_l^{\kappa g}(z) \equiv W_\kappa(z) W_g(z) P(l/\chi(z);z)$ of the theoretical cross-spectrum (Eq.~\ref{eqPS}; shown in Fig.~\ref{figKernels}). At $l = 200$, where the signal-to-noise of the cross-spectrum peaks, the mean redshift of the kernel is $\int z C_{200}^{\kappa g}(z) dz / C_{200}^{\kappa g}= 1.5$. We adopt $z_{\rm eff} = 1.5$ as the effective redshift of the measurement, estimating $b(z_{\rm eff})= 3.5 \pm 0.8$. We note however that we are sensitive to a range of redshifts: at $l = 200$ the kernel is non-negligible ($>10\%$ of its peak value) out to redshifts $z > 3$, and the kernel shifts to higher redshifts at smaller scales. High-redshift sources make an important contribution to the small-scale cross-spectrum. 

We consider a set of variations to the bias model. First we fit the data with a redshift-independent bias model, varying the amplitude $b$. We find $b = 2.8 \pm 0.6$, with $\chi^2_{\rm bf}$ = 11.2. Our data alone cannot distinguish a redshift-independent bias model from an evolving bias model, although our redshift-dependent model is more physically motivated by theoretical and empirical observations \citep[e.g.,][]{Wilman:2008, Lindsay:2014b}.

We probe the flux dependence of the AGN bias by splitting the FIRST sources into two roughly equal-sized subsamples, with a partition at 2.5~mJy. We create new maps of these FIRST sources, as described in Section~\ref{ssFIRST}. From SKADS we estimate the normalised redshift distribution for each subsample, finding they are equal to within $\approx 15\%$ across $0.3 < z < 4$, calculate $z_{\rm eff} = 1.5$ for both subsamples, and estimate the bias amplitude is $\approx 20\%$ higher in the bright sample. We repeat the analysis of Section~\ref{ssAnalysis}, finding $b(z_{\rm eff}; F>2.5~$mJy$) = 4.0 \pm 1.1$ and $b(z_{\rm eff}; F<2.5~$mJy$) = 3.0 \pm 1.1$. This is consistent with the expectation that the high-flux sample selects preferentially for the most-luminous sources, and these sources lie in the most highly-biased environments \citep[e.g.,][]{Mo:1996, Raccanelli:2012}. 

We investigate whether the data provide information about the (largely unknown) high-redshift bias of radio-loud AGN. Here we fix the bias to the fiducial model at redshifts $z<1.5$ and to a redshift-independent value above this. We constrain the high redshift bias to be $b(z>1.5) = 4.1 \pm 1.7$. The detection significance is reduced relative to the full sample because only high-redshift sources ($\approx 1/3$ of the total) constrain this parameter. The increase in bias between low and high redshift samples is consistent with the result of \cite{Lindsay:2014b}, who show that the bias $b(z)$ continues to increase above redshift $z = 2$, although we note the significance is low. 

We divide the source sample into those that have redshift estimates or not. A fraction of $0.27$ of the FIRST sources have a reliable optical match as described in Section~\ref{ssFIRST}. The redshift distribution of these sources is strongly weighted to low redshifts, peaking around $z = 0.5$. Following \cite{Lindsay:2014b} we can estimate the redshift distribution of the remaining sources by comparison with the SKADS simulated radio catalogue used to construct the model redshift distribution for the full sample. We construct independent overdensity maps for these two radio populations (with/without redshift) and recompute the data cross-spectra. We also recalculate the theoretical cross-correlation curves, accounting for the different source distributions, as a function of a redshift-independent bias term $b$. We find $b = 2.1 \pm 1.1$ at an effective redshift $z_{\rm eff} = 0.5$ for the sample with redshifts, and $b = 3.1 \pm 0.8$ at an effective redshift $z_{\rm eff} = 1.6$ for the sample without redshifts, shown in Fig.~\ref{figPosterior}. Although not formally significantly different, this is consistent with an increasing bias as a function of redshift. 
\subsection{Comparison to previous bias measurements}
\cite{Geach:2013} find a constant linear bias $b = 1.61 \pm 0.22$ at an effective redshift $z \approx 1$ for IR-selected quasars from WISE in cross-correlation with the SPT convergence map. At the same effective redshift our bias amplitude determination corresponds to $b(z = 1)= 2.6 \pm 0.6$. Their quasar sample is shallower (42 sources deg$^{-2}$) than in the FIRST maps presented here (71 sources deg$^{-2}$), and the predominant signal comes from $z<2$ sources (there are expected to be no $z > 3$ sources). The higher bias determination presented here is consistent with a more highly biased population of sources being sampled. 

\cite{Sherwin:2012} constrain a constant linear bias $b = 2.5 \pm 0.6$ for optically-selected quasars from SDSS totalling 75 sources deg$^{-2}$. The redshift distribution of these sources peaks at $z = 1.4$. \cite{White:2012} determine $b = 3.8 \pm 0.3$ from the two-point correlation function of quasars in the Baryon Oscillation Spectroscopic Survey across the redshift range $2.2 < z < 2.8$. Comparing with Fig. \ref{figPosterior}, this is in good agreement with our result and assumed bias model. 

\cite{Lindsay:2014b} measure the bias as a function of redshift by auto-correlation of radio sources from the GAMA survey to the same depth (1~mJy) as this analysis. Assuming comoving clustering, their low redshift measurement, $b(z \approx 0.5) = 2.13^{+0.90}_{-0.76}$, is consistent with the results presented here, while at high redshift they find $b(z \approx 1.5) = 9.45^{+0.58}_{-0.67}$, significantly higher than seen in this analysis.

Our result probes the multipole range $100 < l < 3000$, corresponding to physical scales $\approx$~2--60 Mpc at the effective redshift $z_{\rm eff}= 1.5$. As seen in Fig.~\ref{figCOMBdata}, low- and high-redshift sources contribute to the cross-spectrum differently as a function of scale. At the scales probed by the \cite{PlanckXVII} lensing cross-correlation analysis with NVSS radio sources ($l < 400$), $z>1.5$ sources contribute $\sim 1/3$ of the total cross-spectrum, whereas at smaller scales these sources contribute equally alongside the $z<1.5$ sources. By measuring the cross-spectrum across a wide-range of scales one might distinguish between low and high redshift sources. Future high-precision determinations of this cross-spectrum will go further in breaking the degeneracy between source populations and constraining the bias as a function of redshift. 

We can translate the constraint on the AGN bias at redshift $z_{\rm eff} = 1.5$ into an inference on the mass of the halo in which the typical AGN source resides. Using the fitting function of \cite{Tinker:2010}, we find $\log(M_{\rm halo} / M_\odot) = 13.6^{+0.3}_{-0.4}$, assuming that haloes virialize at a density ratio $\Delta = 200$ times that of the universe at the epoch of formation. This observed mass is higher than seen in e.g., \cite{Sherwin:2012}: $\log(M_{\rm halo} / M_\odot) = 12.9^{+0.3}_{-0.5}$; and \cite{Geach:2013}: $\log(M_{\rm halo} / (h^{-1} M_\odot)) = 12.3^{+0.3}_{-0.2}$. This is consistent with the observations of e.g., \cite{Shen:2009} and\cite{Hatch:2014} that the environments of radio-loud AGN are significantly denser than for radio-quiet AGN. 

We find a high bias for these sources compared to optically- and IR-selected AGN. This analysis provides complementary information by probing the bias of radio-selected AGN which, in the context of previous work, is indicative of bias evolution and a very large halo mass for these sources. The broad picture is that of an increasing bias as a function of redshift, and of radio-loud AGN occupying more massive haloes than radio-quiet AGN across a similar redshift range. 

Our findings are in line with studies of the stellar masses \citep[e.g.,][]{Jarvis:2001, Seymour:2007} and environments \citep[e.g.,][]{Wylezalek:2013, Hatch:2014} of powerful radio sources to high redshift. Specifically, we find strong evidence that powerful radio sources are more highly biased tracers of the dark matter density field than other AGN that are detectable to high redshift  \citep[e.g., quasars;][]{Sherwin:2012, Geach:2013}. As well as being important for tracing the underlying dark matter distribution with techniques such as described in \cite{Ferramacho:2014}, this also suggests that mechanical feedback from the jets of {\em powerful} radio AGN, should only have a significant effect on the level of star formation within the most massive dark matter haloes at all epochs. However, we note that such an effect can not only have an impact on both the AGN host galaxy \citep[e.g.,][]{Croton:2006, Bower:2006, Hopkins:2006, Dubois:2013, Mocz:2013}, but also the wider cluster environment \citep[e.g.,][]{Rawlings:2004b}.
\subsection{Modelling limitations and astrophysical systematics}
\label{ssLimitations}
The SKADS simulation is populated using empirical radio luminosity functions as described in \cite{Wilman:2008}. Extrapolation of the empirical luminosity functions into unobserved regimes will lead to inaccuracies in the inferred redshift distribution and bias model. To investigate the sensitivity of our measurement to uncertainties about the source redshift distribution, we recompute the theoretical spectra, unrealistically removing all sources above redshift $z > 3$ when calculating $dn/dz$; at high redshift the underlying $dn/dz$ is most uncertain and likely depends on radio luminosity \citep[e.g.,][]{Jarvis:2000, Jarvis:2001, Wall:2005, Rigby:2011}. Fitting the theoretical cross-spectrum as in Section~\ref{sResults}, we find $b(z_{\rm eff} = 1.2)= 3.2 \pm 0.8$, representing a $\sim0.25\sigma $ shift from the primary result under this significant perturbation of the theoretical redshift distribution. We thus do not expect that the source distribution uncertainty strongly biases our result, although future analyses with higher statistical power will require careful consideration of this systematic uncertainty. 

We fix the cosmology to the {\sl Planck} best-fit values throughout this analysis, which could affect the inference of the AGN bias. However, the significant ($40\sigma$) detection of the {\sl Planck} lensing auto-spectrum means that model uncertainty from the cosmology is sub-dominant with respect to astrophysical uncertainties \citep{Planck2015XV}. Perturbing the {\sl Planck} best-fit cosmological parameters by $+1\sigma$ and recomputing the theoretical cross-spectrum, $C_l^{\kappa g}$, the amplitude is shifted by $<6\%$ across all relevant scales; we thus neglect this source of systematic uncertainty. 

Potential astrophysical systematic contaminants include infrared sources, Sunyaev-Zeldovich (SZ) clusters and Galactic cirrus. \cite{Sherwin:2012} show that these constitute small effects on the measured cross-spectrum between quasars and lensing ($<10\%$ in total), negligible at the level of statistical uncertainty in this analysis. Although the sources studied in \cite{Sherwin:2012} are optically-selected AGN, we expect the result to hold for the radio-loud AGN of this analysis given the roughly similar redshift distributions. Furthermore, bright radio sources ($\gtrsim 5$~mJy) in the CMB temperature and polarization maps are subtracted prior to lensing reconstruction, using a match-filtered source template map, thus mitigating radio-source contamination in the CMB convergence map \citep{Das:2011,vanEngelen:2014}. 
\section{Conclusions}
\label{sConclusion}
We present a measurement of the angular cross-power spectrum between lensing convergence from ACT and the overdensity of radio sources identified in the FIRST survey, rejecting the null-hypothesis of no correlation at $4.4\sigma$ significance. The data are well fit by the {\sl Planck} best-fit $\Lambda$CDM cosmological model where we model the source population with a redshift-dependent bias. We interpret the result in terms of a constraint on the bias of AGN, which dominate the FIRST sample, considering various bias models and data splits to probe different redshift regimes and AGN populations, and put these in the context of previous measurements of AGN bias. We translate the bias determination into a constraint on the mass of the host haloes, corroborating previous work showing that the environments of radio-loud AGN are more dense than those of optically-selected AGN. 

We consider various sources of systematic uncertainty, both astrophysical contaminants and modelling limitations. We conclude that our results are robust to these effects. As deeper and wider radio surveys and improved lensing maps become available, these systematic effects will become increasingly important to measure and model accurately. The auto- and cross-spectra $\{ C_l^{gg}, C_l^{\kappa \kappa}, C_l^{\kappa g} \}$ provide complementary information about the large-scale structure they probe, with the cross-spectrum in particular being robust to systematic biases particular to each dataset. A full analysis will simultaneously estimate the three power spectra, marginalizing over uncertainty in the redshift distribution and cosmology \citep{Pearson:2014}. With current data there are strong degeneracies in the cross-spectrum amplitude between sources from different redshifts. The shape of the power-spectra contain information about the bias evolution, and larger, more sensitive surveys will allow us to break these degeneracies. 

The measurement of the high bias (and correspondingly large halo mass) of this radio population, relative to other dark matter tracers, indicates that these sources would be useful in the multi-tracer technique of \cite{Ferramacho:2014}. Using all the information in auto- and cross-correlations between multiple tracers, which differentially trace the dark matter, will provide tight constraints on primordial non-Gaussianity by reducing the impact of cosmic variance at large scales. 

The SKA will serve as a deep probe of large-scale structure in the universe, it will be limited by different systematics than optical surveys, and the observed source distribution will be skewed to higher redshifts than either LSST or Euclid \citep{Jarvis:2015}. \cite{Kirk:2015} show that next-generation CMB lensing experiments, in combination with the SKA, will constrain the amplitude of the lensing-radio density cross-spectrum to the sub-percent level. With tight constraints on cosmology, this translates into $<1\%$ uncertainty on the bias amplitude, offering broad scope for probing the history and evolution of AGN. Future high-precision measurements of $C_l^{\kappa g}$ will use information about the shape of the cross-spectrum, and source tomography, to constrain the bias as a function of redshift, calibrating galaxy redshift surveys and constraining extensions to $\Lambda$CDM.

The cross-correlation of CMB lensing with tracers of large-scale structure will become an increasingly important calibrator for future high-precision galaxy and weak-lensing surveys. 
\section*{Acknowledgments}
RA is supported by an STFC Ph.D. studentship. This work was supported by the U.S. National Science Foundation through awards AST-0408698 and AST- 0965625 for the ACT project, as well as awards PHY-0855887 and PHY-1214379. Funding was also provided by Princeton University, the University of Pennsylvania, Cornell University, and a Canada Foundation for Innovation (CFI) award to UBC. ACT operates in the Parque Astron\'o{}mico Atacama in northern Chile under the auspices of the Comisi\'o{}n Nacional de Investigaci\'o{}n Cient\'i{}fica y Tecnol\'o{}gica de Chile (CONICYT). Computations were performed on the GPC supercomputer at the SciNet HPC Consortium. SciNet is funded by the CFI under the auspices of Compute Canada, the Government of Ontario, the Ontario Research Fund Ð Research Excellence; and the University of Toronto. The development of multichroic detectors and lenses was supported by NASA grants NNX13AE56G and NNX14AB58G. Funding from ERC grant 259505 supports SN, JD, and TL. RD was supported by CONICYT grants QUIMAL-120001 and FONDECYT-1141113. We gratefully acknowledge support from the Misrahi and Wilkinson research funds.

\bibliographystyle{mn2e}
\bibliography{ref}
\label{lastpage}
\clearpage
\end{document}